\documentclass[aps,pre,preprint,showpacs]{revtex4}

\usepackage{graphicx}
\usepackage{dcolumn}
\usepackage{bm}

\begin{document}
\title{Diffusion of Colloidal Fluids in Random Porous Media}

\author{M. A. Ch\'avez-Rojo}

\affiliation{Facultad de Ciencias Qu\'{\i}micas, Universidad
Aut\'onoma de Chihuahua, Venustiano Carranza S/N, 31000 Chihuahua,
Chih., M\'exico. }
\author{R. Ju\'arez-Maldonado and M. Medina-Noyola}

\affiliation{Instituto de F\'{\i}sica ``Manuel Sandoval Vallarta'',
Universidad Aut\'onoma de San Luis
Potos\'{\i}, Alvaro Obreg\'on 64, 78000 San Luis Potos\'{\i}, S.L.P., M\'exico.}

\date{\today}

\begin{abstract}

The diffusive relaxation of a colloidal fluid adsorbed in a porous
medium depends on many factors, including the concentration and
composition of the adsorbed colloidal fluid, the average structure
of the porous matrix, and the nature of the colloid-colloid and
colloid-substrate interactions. A simple manner to describe these
effects is to model the porous medium as a set of spherical
particles fixed in space at random positions with prescribed
statistical structural properties. Within this model one may
describe the relaxation of concentration fluctuations of the
adsorbed fluid by simply setting to zero the short-time mobility of
one species (the porous matrix) in a theory of the dynamics of
equilibrium colloidal mixtures, or by extending such dynamic theory
to explicitly consider the porous matrix as a random external field,
as recently done in the framework of mode coupling theory [V.
Krakoviack, Phys. Rev. Lett. {\bf 94}, 065703 (2005)]. Here we
consider the first approach and employ the self-consistent
generalized Langevin equation (SCGLE) theory of the dynamics of
equilibrium colloidal mixtures, to describe the dynamics of the
mobile component. We focus on the short- and intermediate-time
regimes, which we compare with Brownian dynamics simulations
involving a binary mixture with screened Coulomb interactions for
two models of the average static structure of the matrix: a porous
matrix constructed by quenching configurations of an equilibrium
mixture in which both species were first equilibrated together, and
a pre-existing matrix with prescribed average structure, in which we
later add the mobile species. We conclude that in both cases, if the
correct static structure factors are provided as input, the SCGLE
theory correctly predicts the main features of the dynamics of the
permeating fluid.

\end{abstract}
\pacs{05.40.-a, 82.70.Dd, 02.70.-c}

\maketitle

Many relevant systems and processes in industry and nature involve
the diffusion of colloidal dispersions through porous media
\cite{sahimi}. The most relevant issues involve either the
equilibrium and phase behavior \cite{lowen}, or the transport and
dynamic properties \cite{sahimi2}, of the permeating colloidal
liquid. Thus, one would like to understand, from a fundamental
perspective, how these properties depend on factors such as the
porosity and morphology of the matrix, the nature of the interaction
forces (among the particles and with the porous matrix), or the
concentration of the colloidal dispersion. Some of these issues
require the development or extension of the fundamental description
of the dynamic behavior of bulk colloidal systems \cite{PUSEY,
NAGELE} to the case in which these systems permeate a porous matrix.

The fundamental study of the phenomena above rely on simplified
models. Thus, a porous medium is sometimes modeled as a simple
geometry (planar slit, cylindrical pore, etc.) to describe local
phenomena, whereas random arrays of locally regular pores
incorporate the intrinsic randomness of most natural or synthetic
porous materials \cite{sahimi}. One may adopt, instead, a simplified
model of a random porous medium, namely, a matrix of spherical
particles with random but fixed positions. This matrix is permeated
by a colloidal liquid, whose dynamics we wish to understand. Such
model systems have been employed to describe mostly equilibrium
structural properties \cite{pizio1}, although simple model
experimental realizations of this system have been prepared
\cite{q2dporous}, in which one could also measure the dynamic
properties of the mobile species. The interpretation of such
measurements requires sound theoretical schemes to describe the
dynamics of the permeating dispersion. One possible approach uses
available theories of the dynamic properties of bulk colloidal
mixtures \cite{nagele2, marco2, todos2} in which the mobility of one
of the species is artificially set equal to zero. Another
possibility is to first reformulate these theories to explicitly
consider the porous matrix as a random external field
\cite{krakoviack1}. In this letter we demonstrate that the first of
these approaches suffices to correctly predict the main features of
the dynamics of the permeating fluid, provided the correct average
static structure of the matrix and of the adsorbed fluid is
available.

This conclusion is based on the use of the multi-component
self-consistent generalized Langevin equation (SCGLE) theory of
colloid dynamics \cite{marco2}, to describe the relaxation of
concentration fluctuations of the mobile component in the model
porous matrix. By setting the free-diffusion coefficient of one
species to zero, this theory is readily adapted to the description
of the dynamics of the model system above, thus allowing the
numerical calculation of the partial intermediate scattering
function $F(k,t)$ of the mobile species. As a concrete and
illustrative application, here we report the theoretical predictions
for a binary colloidal mixture of particles interacting through
screened Coulomb potentials in which one of the two species plays
the role of the porous matrix. The theoretical results for this
specific system are compared with the corresponding results of a
Brownian dynamics (BD) simulation on a model system consisting of
$N= N_1 + N_2$ Brownian particles in a volume $V$, with $N_1(=n_1V)$
particles of species 1 and $N_2(=n_2V)$ particles of species 2
interacting through direct forces, but not through hydrodynamic
interactions.

We carry out two kinds of computer experiments that differ in the
manner we generate the structure of the porous matrix. In the first
kind, we let the $N$ particles of both species to undergo Brownian
motion, according to the conventional Brownian dynamics algorithm
\cite{TILDES} with the same free-diffusion coefficient $D^0_1 =
D^0_2 =D^0$ until equilibrium is reached. At this point, we
artificially arrest the motion of the particles of species 2 by
setting $D^0_2=0$ at an arbitrary configuration. In the second kind
of experiments, a pre-existing matrix is formed in the absence of
the mobile species, by choosing the arrested configurations
according to a prescribed distribution, afterward ``pouring" the
mobile particles into this matrix of obstacles. The prescribed
average structure of the matrix that we consider below corresponds
to the structure of an equilibrium mono-component fluid of species
2. In both cases, after choosing a particular configuration of the
matrix, we let the mobile species equilibrate in the external field
of the fixed particles at that particular frozen configuration, and
then proceed to the calculation of the dynamic properties of
interest. Since these properties depend on the specific
configuration of the fixed particles, and  in order to average out
this dependence, the results presented here correspond to an average
over more than 50 different configurations of the porous matrix. In
both cases we also record the radial distribution functions between
the two species, to be employed as the static input required by the
SCGLE theory. In our illustrative application, the direct
interactions are described by a hard-sphere plus a strong repulsive
screened Coulomb interaction. For simplicity, we assume that both
species have the same hard-sphere diameter $\sigma$, so that the
potential (in units of the thermal energy $k_BT = \beta^{-1}$) is
given by $\beta u_{\alpha \beta}(r) =+\infty$ for $r<\sigma$, and
for $ r
> \sigma$ by

\begin{equation}
\beta u_{\alpha \beta}(r)  = \sqrt{K_{\alpha } K_{ \beta}}
\frac{\exp\left[
-z(r/\sigma -1)\right] }{r/ \sigma}\ . \label{dlvo0}%
\end{equation}
One may think of the parameter $\sqrt{K_{\alpha}}$ as proportional
to the electric charge $Q_{\alpha}$ of species $\alpha$, and the
parameter $z=\kappa \sigma$ is the dimensionless inverse Debye
length $\kappa$ \cite{NAGELE}.

The relevant dynamic information of an equilibrium $\nu$-component
colloidal suspension is contained in the $\nu \times \nu$ matrix
$F(k,t)$ whose elements are the {\it partial intermediate scattering
functions} $F_{\alpha \beta}(k,t)\equiv \left\langle n_{\alpha}({\bf
k},t) n_{\beta}(-{\bf k}^{\prime},0)\right\rangle$ where
$n_{\alpha}({\bf k},t)\equiv \sum_{i=1}^{N_{\alpha}} \exp[i{\bf
k}\cdot {\bf r}_i(t)]/\sqrt{N_\alpha}$, with ${\bf r}_i(t)$ being
the position of particle $i$ of species $\alpha$ at time $t$. The
initial value $F_{\alpha \beta}(k,0)$ is the partial static
structure factor $S_{\alpha \beta}(k)$ \cite{HANSEN, NAGELE}. In our
simulation experiment, we are interested in the dynamic properties
of the mobile species, represented by $F(k,t)\equiv F_{11}(k,t)$.
This dynamic property will be theoretically calculated applying the
multi-component self-consistent generalized Langevin equation
(SCGLE) theory of colloid dynamics \cite{marco2} with the particular
condition $\nu=2$ and $D^0_2=0$.

The SCGLE theory, explained in more detail in Refs.\ \cite{marco2,
todos2}, is summarized by a self-consistent system of equations for
the $\nu \times \nu$ matrices $F(k,t)$ and $F^{(s)}(k,t)$ (the
latter defined as $F_{\alpha\beta}^{(s)}(k,t)\equiv \delta_{\alpha
\beta}\left\langle \exp {[i{\bf k}\cdot \Delta{\bf R^{(\alpha
)}}(t)]} \right\rangle $, where $\Delta{\bf R}^{(\alpha )}(t)$ is
the displacement of any of the $N_{\alpha }$ particles of species
${\alpha}$ over a time $t$, and $\delta_{\alpha \beta}$ is
Kronecker's delta function. Written in matrix form and in Laplace
space, and omitting the explicit $k$-dependence, the self-consistent
system of equations reads

\begin{equation}\label{fdz}
F(z)=\left\{z+(I+[\Delta\zeta^*(z)]\lambda)^{-1}k^{2}DS^{-1}\right\}^{-1}S,
\end{equation}
and

\begin{equation}\label{fsdz}
F^{(s)}(z)=\left\{z+(I+[\Delta\zeta^*(z)]\lambda)^{-1}k^{2}D\right\}^{-1},
\end{equation}
where $S$ is the matrix of partial static structure factors, $D$ and
$\lambda(k)$ are diagonal matrices given by $D_{\alpha \beta} \equiv
\delta_{\alpha \beta} D^0_{\alpha}$ and $\lambda_{\alpha\beta} (k) =
\delta_{\alpha \beta} [1+(k/k^{(\alpha)}_c)^2]^{-1}$, where
$k^{(\alpha)}_c$ is the location of the first minimum (following the
main peak) of $S_{\alpha \alpha}(k)$. $\Delta\zeta^*(t)$ is a
diagonal matrix with its diagonal element $\Delta\zeta_\alpha^*(t)$
given by

\begin{equation}
\Delta \zeta ^{*} _{\alpha}(t) =\frac{D^0_{\alpha}} {24\pi^3}\int
d^3k k^2 [F^{(s)}(t)]_{\alpha\alpha} [c \sqrt{n} F(t) S^{-1}
\sqrt{n} h]_{\alpha\alpha}, \label{5}
\end{equation}
where the elements of the $k$-dependent matrices $h$ and $c$ are the
Fourier transforms $h_{\alpha\beta}(k)$ and $c_{\alpha\beta}(k)$ of
the Ornstein-Zernike total and direct correlation functions,
respectively. Thus, $h$ and $c$ are related to $S$ by $S =
I+\sqrt{n}h\sqrt{n} = [I-\sqrt{n}c\sqrt{n}]^{-1}$, with  the matrix
$\sqrt{n}$ defined as $[\sqrt{n}]_{\alpha\beta} \equiv
\delta_{\alpha\beta}\sqrt{n_\alpha}$.

\begin{figure}
\includegraphics[scale=0.35]{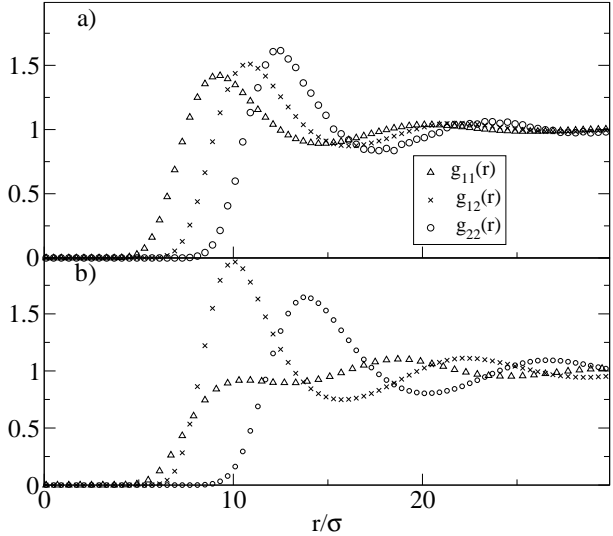}
\caption{\label{fig:static} Brownian dynamics simulated radial
distribution functions $g_{\alpha\beta}(r)$ of a colloidal fluid
(species 1) diffusing through a porous matrix formed by a second
species of fixed particles, interacting with the repulsive Yukawa
potential with fixed screening parameter $z=0.15$, volume fractions
$\phi_1= \phi_2=2.2\times 10^{-4}$ and repulsion strength parameters
$K_1=100$ and $K_2=500$ with the porous matrix generated in the
presence (a), and in the absence (b), of the mobile species. }
\end{figure}

The control parameters of our system are the interaction parameters
$z$, $K_1$ and $K_2$, and the volume fractions $\phi_1$ and $\phi_2$
(with $\phi_{\alpha}\equiv \pi n_{\alpha}\sigma^3/6$). Here we fix
the value of the screening parameter to $z=0.15$, and start by
considering an equi-molar mixture with $\phi_1=\phi_2=2.2\times
10^{-4}$. In our first simulation experiment we start with the
simplest case, namely, a mono-disperse suspension of $N=N_1 + N_2$
identical particles interacting with the same pair potential
($K_1=K_2=100$), which execute Brownian motion. After
thermalization, we stop the motion of half of them, and let the
other half constitute the mobile species. Two additional experiments
of the same kind were performed for systems with the same parameters
as above, but varying the coupling parameters $K_1$ and $K_2$. Thus,
the second experiment corresponds to a more interacting system,
$K_1=K_2= 500$, and the third to an asymmetric mixture such that the
matrix is formed by the more strongly interacting particles, $K_2=
500$ and $K_1= 100$. These three experiments belong to the first
kind referred to above, i.e., they involve a matrix whose average
static structure is identical to the partial static structure factor
$S_{22}(k)$ of an equilibrium mixture of both species. In this kind
of experiments, $S_{22}(k)$ and the other static structural
properties may be determined during the initial equilibration stage,
before arresting the motion of the matrix. As an example, in Fig.\ 
\ref{fig:static}.a we present the various radial distribution
functions $g_{\alpha\beta}(r)$ simulated in this manner in the third
of these experiments.

We also performed parallel experiments of the second kind, involving
a pre-existing matrix with prescribed average structure. The
prescribed structure we chose corresponds to the equilibrium
structure of a \emph{mono-component} system containing only species
2. Thus, we first let the $N_2$ particles equilibrate, and then
freeze an arbitrary configuration in which we then place the $N_1$
particles of the other species. The static and the dynamic
properties involving the mobile species are then simulated after the
fluid of species 1 equilibrates in the external field of the matrix
in that particular configuration. The results are then averaged over
a sufficient number ($\sim$50) of configurations of the matrix. In
Fig.\ \ref{fig:static}.b we present the resulting
$g_{\alpha\beta}(r)$ corresponding to the third experiment of this
second kind, involving a system with the same parameters as in Fig.\ 
\ref{fig:static}.a.

\begin{figure}
\includegraphics[scale=0.35]{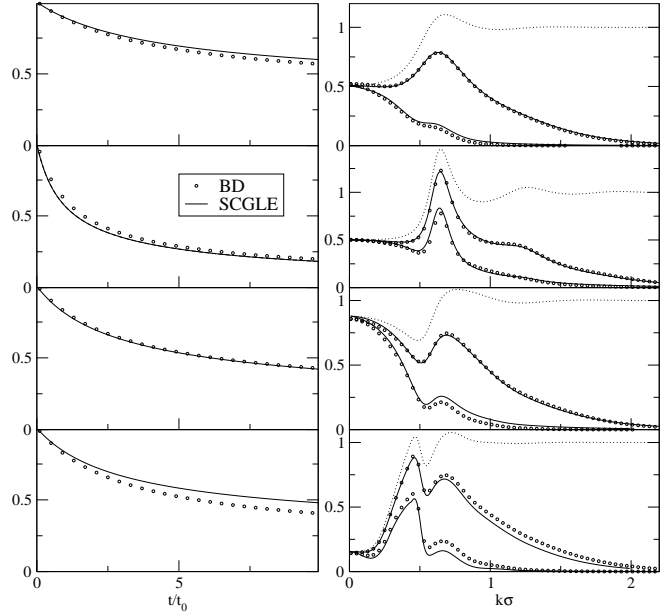}
\caption{\label{fig:dynamic} Time-dependent self diffusion
coefficient $D^*(t)$ (left column) and partial intermediate
scattering function $F(k,t)$ for $t=0, \ t_0$, and $10t_0$ (right
column) of the diffusive species permeating the porous matrix,
interacting with the repulsive Yukawa potential with fixed screening
parameter $z=0.15$ and volume fractions $\phi_1= \phi_2=2.2\times
10^{-4}$, but with parameters $K_1$ and $K_2$ given by $K_1=K_2=100$
(first row), $K_1=K_2=500$ (second row) and $K_1=100$ and $K_2=500$
(third and fourth rows). The symbols represent Brownian dynamics
results and the solid lines are the SCGLE theoretical predictions.}
\end{figure}

These simulated structural properties are now employed as the static
input needed by the SCGLE theory, thus avoiding the use of liquid
state approximations \cite{HANSEN}. The predictions of the SCGLE
theory for the dynamic properties of the Brownian fluid immersed in
the porous matrix are presented in Fig.\ \ref{fig:dynamic} for the
three experiments of the first kind and for only the last experiment
of the second kind, namely, that involving the asymmetric mixture
with $K_2= 500$ and $K_1= 100$. These results are compared with the
corresponding BD results for the normalized time-dependent diffusion
coefficient $D^*(t) \equiv <(\Delta {\bf r}(t))^2>/(6D^0t)$ and for
the collective intermediate scattering function $F(k,t) =
F_{11}(k,t)$ of the mobile species at fixed $k$ and for the times
$t=0,\ t_0$, and $10t_0$, with $t_0 \equiv \sigma^2/D^0$. Notice
that for $t=10t_o$, $D^*(t)$ has relaxed from its initial value of 1
towards, and close to, its asymptotic value, characteristic of the
long-time regime. Thus, the illustrative data in this figure covers
the so-called short- and intermediate-time regimes, most easily
accessible by Brownian dynamics simulations or video-microscopy
experiments.

From information such as that summarized in Fig.\ \ref{fig:dynamic}
we may highlight the following. First, the description of the SCGLE
theory for symmetric systems (illustrated by the first two rows) is
highly accurate in the time-regimes illustrated in the figure.
Second, the comparisons in Fig.\ \ref{fig:dynamic} have essentially
the same quality as the corresponding comparisons involving fully
thermalized mixtures in which both species diffuse \cite{marco2}.
Third, the theoretical predictions for $F(k,t)$ in asymmetric
systems for both kinds of experiments (last two rows) have similar
levels of quantitative accuracy; the largest discrepancies with the
simulation data are illustrated by the results of the fourth
experiment, without being particularly severe. Clearly, any
improvement will involve either reformulating the intrinsic
approximations of the SCGLE theory or developing an extension
similar to that carried out by Krakoviack \cite{krakoviack1} in the
framework of mode coupling theory (MCT) \cite{goetze1}. At best,
however, such improvements will only add to the quantitative
accuracy of the present results and will be of the magnitude
illustrated in Fig.\ \ref{fig:dynamic}.

Thus, the comparisons above indicate that the SCGLE theory, devised
to describe the dynamics of equilibrium colloidal mixtures, provides
a useful approach to the dynamics of a mono-disperse suspension
permeating a porous medium formed by a random array of other
colloidal particles. This approach may now be applied to explore
other interesting phenomena such as, for example, transitions of
dynamic arrest of colloidal {\em mixtures} in porous media
\cite{lowen}. In fact, the present theory, complemented by adequate
liquid state approximations for the static structure \cite{HANSEN},
may be used as a fully theoretical first-principles approach to
qualitatively scan other regions of the state space to locate
interesting dynamic phenomena for which simulations or experiments
would be difficult to carry out, or are not yet available. In doing
this we only loose quantitative precision, but no qualitative
accuracy. In fact, one of the objectives of adapting the SCGLE
theory to the description of the dynamics of colloidal dispersions
adsorbed in porous media is the description of dynamic arrest
phenomena in these systems. The analysis of its accuracy in the
short- and intermediate times presented here is an important step in
the process of assessing its overall reliability. We mention,
however, that the SCGLE theory has been successfully employed to
describe dynamic arrest in bulk mono-disperse \cite{todos2, rmf,
todos1} and multi-component \cite{rigo1} colloidal systems. In fact,
we can communicate that the SCGLE theory presented here also
predicts dynamic arrest in the systems considered above upon
increasing the coupling parameters $K_1$ and/or $K_2$ or by varying
other control parameters of the system. In this manner, we can
outline the full dynamic arrest phase diagram of this model system.
In particular we can report that for systems involving only hard
sphere interactions, and  under the specific conditions studied by
Krakoviack using MCT \cite{krakoviack1}, we obtain essentially the
same dynamic arrest scenario that he derives from MCT. Our approach,
however, opens the possibility to study more complex situations in a
simpler manner and from an alternative perspective to that provided
by MCT. The details of these specific advances, however, will be
reported separately.

\begin{center}
{\bf ACKNOWLEDGMENTS}
\end{center}

This work was supported by the Consejo Nacional de Ciencia y
Tecnolog\'{\i}a (CONACYT, M\'exico), through grants No. 2004-C01-47611
and No. 2006-C01-60064, and by FAI-UASLP. The authors are grateful to
Profs. J. Bergenholtz, A. Banchio, G. N\"agele, and M.
Ch\'avez-P\'aez for useful discussions.

\end{document}